\documentclass[preprint]{aastex}

\RequirePackage{color}

\newcommand{\gsim}{\hspace{0.3em}\raisebox{0.4ex}{$>$}\hspace{-0.75em}\raisebox{-.7ex}{$\sim$}\hspace{0.3em}}
\newcommand{\lsim}{\hspace{0.3em}\raisebox{0.4ex}{$<$}\hspace{-0.75em}\raisebox{-.7ex}{$\sim$}\hspace{0.3em}}

\shorttitle{Evolution of dwarf satellites}
\shortauthors{Okayasu \& Chiba}

\begin{document}

\title{Evolution of dwarf spheroidal satellites in the common surface-density dark halos}
\author{Yusuke~Okayasu\altaffilmark{1} and Masashi~Chiba\altaffilmark{1}}
\altaffiltext{1}{Astronomical Institute, Tohoku University, Aoba-ku, Sendai 980-8578, Japan}
\email{y.okayasu@astr.tohoku.ac.jp; chiba@astr.tohoku.ac.jp}

\begin{abstract}
We investigate the growth histories of dark matter halos associated with dwarf satellites
in Local Group galaxies and the resultant evolution of the baryonic component. 
Our model is based on the recently proposed property that the mean surface density of a 
dark halo inside a radius at maximum circular velocity $V_{\rm max}$ is universal 
over a large range of $V_{\rm max}$. Following that this surface density of
20 $M_{\odot}$~pc$^{-2}$ well explains dwarf satellites in the Milky Way and Andromeda,
we find that the evolution of the dark halo in this common surface-density scale
is characterized by the rapid increase of the halo mass assembled by the redshift
$z_{\rm TT}$ of the tidal truncation by its host halo, at early epochs of
$z_{\rm TT} \gsim 6$ or $V_{\rm max} \lsim 22$ km~s$^{-1}$.
This mass growth of the halo is slow at lower $z_{\rm TT}$ or larger $V_{\rm max}$.
Taking into account the baryon content in this dark halo evolution, under the
influence of the ionizing background radiation,
we find that the dwarf satellites are divided into roughly two families:
those with $V_{\rm max} \lsim 22$ km~s$^{-1}$ having high star formation
efficiency and those with larger $V_{\rm max}$ having less efficient star formation.
This semi-analytical model is in agreement with the high-resolution numerical
simulation for galaxy formation and with the observed star formation histories for
Fornax and Leo~II. This suggests that the evolution of a dark halo may play
a key role in understanding star formation histories in dwarf satellites.
\end{abstract}


\section{Introduction}
Standard $\Lambda$ cold dark matter ($\Lambda$CDM) theory is successful for 
explaining the large-scale structure of the Universe, including the temperature 
fluctuations of the cosmic microwave background and the spatial distributions of galaxies 
and their clusters, on scales lager than about 1~Mpc \citep{Tegmark2004}. However, our 
understanding of structure formation is still incomplete on scales smaller than 1~Mpc,
i.e. scales of galaxies like our own and their satellite galaxies, which are basic luminous
parts of the Universe.

The tension between the theory and observation 
on small scales includes the missing satellite problem \citep{Klypin1999,Moore1999},
too-big-to-fail problem \citep{Boylan-Kolchin2011,Boylan-Kolchin2012}, and
the problems for explaining cored central densities
\citep[e.g.,][]{Moore1994,Burkert1995,de Blok2001,Swaters2003,Gilmore2007,Oh2011} and 
anisotropic spatial distribution of dwarf satellites 
\citep[e.g.,][]{Kroupa2005,McConnachie2006,Ibata2013,Pawlowski2012,Pawlowski2013,Pawlowski2015}.
Solutions to these small-scale issues in $\Lambda$CDM theory have been considered
from two different aspects.
One is to modify the standard theory of dark matter on small scales, such as theories
based on warm or self-interacting dark matter \citep[e.g.,][]{Maccio2010,Lovell2012,Vogelsberger2012}. 
Another way to solve the issues is to rely on the role of baryonic physics in the formation 
of galaxies and their satellites, including the suppression effects of
the ionizing UV background and stellar feedback such as supernova explosion
on galaxy formation as well as tidal effects of a centrally concentrated host galaxy
on its satellites
\citep[e.g.,][]{Di_Cinto2014,Madau2014,Pawlowski2015,Zhu2016,Sawala2016}.

In this context, nearby dwarf spheroidal (dSph) galaxies as satellites of the Milky Way 
and Andromeda galaxies provide ideal sites for studying the nature of dark matter, 
because stellar dynamics of the resolved member stars suggest that these galaxies are 
significantly dominated by associated dark matter halos with mass-to-light ratios of 10 to 1000 
or more \citep{Gilmore2007,Simon2007}. For this reason, extensive 
dynamical analyses for these local dSphs have been performed to set constraints on the 
detailed internal structures of their dark halos and ultimately to get important insights 
into the nature of dark matter on small scales.

These dynamical analyses of dSphs revealed several universal structures of their dark 
halos. \citet{Mateo1993} and more recently, \citet{Strigari2008}, proposed,
based on spherically symmetric mass models, that all the dSphs 
with luminosities in the range of $10^3$ to $10^7$ $L_{\odot}$ have a common 
dark-halo mass of $M_{300} \sim 10^7$ $M_{\odot}$ inside an inner 300~pc radius. 
Following this work, \citet{Maccio2009} showed that this constancy of $M_{300}$ can 
be naturally explained in the framework of $\Lambda$CDM theory combined with 
galaxy formation physics, where the latter process narrows an allowed range of circular 
velocities of dark halos. More recently \citet{Milosavljevic2014} showed
the growth history of dark halos in this common $M_{300}$ scale and the 
resultant impacts on the star formation activity in dwarf satellite galaxies.

Other work including \citet{Kormendy2004}, \citet{Donato2009} and \citet{Gentile2009}
have shown that for the assumed cored isothermal profile or so-called Burkert profile
\citep{Burkert1995} with a density $\rho_0$ and radius $r_0$ of a core component,
the central surface density of a cored dark halo, $\mu_{0D} = \rho_0 r_0$,
is found to be constant, irrespective of the observed $B$-band luminosities of 
galaxies. The similar universality of a central surface density of a cored dark halo has been 
proposed by \citet{Salucci2012}, arriving at $\mu_{0D} \simeq 
140$ $M_{\odot}$~pc$^{-2}$ for both cored isothermal and Burkert profiles. These work 
indeed provide important constraints on the structure of dark halos and thus on the 
nature of dark matter.

Recently, \citet{Hayashi2015a}, hereafter referred to as HC15a, proposed a yet another 
universal scale for the internal structure of dark halos, motivated by the facts that 
realistic dynamical analysis of non-spherical dSphs based on axisymmetric Jeans 
equations yields a rather large dispersion in the values of $M_{300}$ \citep{Hayashi2012}
and that the constancy of $\mu_{0D}$ cannot be applicable to any density 
profiles including a dark halo with a central cusp. HC15a showed for galaxies spanning the 
circular velocities of 10 to 400 km~s$^{-1}$ that a mean surface density of a dark halo 
within the radius of the maximum circular velocity, hereafter denoted as 
$\Sigma_{V_{\rm max}}$, is remarkably constant, irrespective of the different density 
distributions in each of the galaxies. It is also shown that this constancy of 
$\Sigma_{V_{\rm max}}$ is applied to all the galaxies with $B$-band luminosities over 
almost 14 orders of magnitude, $M_B = -8$ to $-22$ mag \citep{Hayashi2015b}, i.e. 
including faint dSphs to bright spiral/elliptical galaxies.

In this paper, we investigate the evolution of dark matter halos in this common 
surface-density scale and the resultant impacts on star formation activity in dwarf 
satellite galaxies, following the work of \citet{Milosavljevic2014} in which the 
common mass scale of $M_{300} = const.$ for dark halos is assumed. We thus show how 
the growth histories of dark halos differ between the case of 
$\Sigma_{V_{\rm max}} = const.$ and that of $M_{300} = const.$ and how these 
different assembly histories of dark matter affect the baryonic content and star 
formation rate in these small-scale dark halos and how the results are compatible with 
the observed properties of dwarf satellite galaxies in the Milky Way and Andromeda 
galaxies.

The paper is organized as follows. In Section 2, we calculate the growth histories of dark 
halos under the constraint of $\Sigma_{V_{\rm max}} = const.$ and the difference 
from $M_{300} = const.$. In Section 3, we consider the baryon content in 
such dark halos and compare with the remained stellar mass in the currently observed 
dSphs to estimate the star formation efficiency. In Section 4, to get further insights into 
the models we develop, we compare with the results of the recent high-resolution 
numerical simulation, so-called the Illustris simulation, and also compare with the star 
formation histories of dSphs derived from the recent observations. Section 5 is devoted 
to our conclusions in this work.
In what follows, we adopt the cosmological parameters of a mean matter density
$\Omega_m=0.2726$, baryon density $\Omega_b=0.0456$ and Hubble parameter $h=0.704$.

\section{Structure and evolution of dark halos in dwarf satellites}
\subsection{The common surface-density scale for Local Group dwarf galaxies}
Our mass model for dark halos in Local Group dSphs is assumed to hold a common 
mean surface density within a radius at maximum circular velocity, $\Sigma_{V_{\rm 
max}} = const.$, proposed by HC15a. In the limit of spherical symmetry, 
$\Sigma_{V_{\rm max}}$ is given as
\begin{equation}
\Sigma_{V_{\rm max}} = \frac{M(r_{\rm max})}{\pi r^2_{\rm max}} \ ,
\label{eq: Sigma}
\end{equation}
where $r_{\rm max}$ is a radius at maximum circular velocity (denoted as $V_{\rm 
max}$) for a given mass profile of a dark halo and $M(r_{\rm max})$ is a mass enclosed 
within $r_{\rm max}$,
\begin{equation}
M(r_{\rm max}) = \int^{r_{\rm max}}_0 4\pi\rho(r')r'^2 dr' \ ,
\label{eq: Mass}
\end{equation}
where $\rho(r)$ denotes a mass density profile of a dark halo. This surface density 
defined in eq. (\ref{eq: Sigma}) is basically proportional to the product of a central 
density, $\rho_{\ast}$, and a scale length, $r_{\ast}$, for any density profile of
a dark halo, such as the so-called Navarro-Frenk-White (NFW) profile \citep{Navarro1996}
and the Burkert profile \citep{Burkert1995}, where the definition of $\rho_{\ast}$
and $r_{\ast}$ depends on a specified form of a given density profile.

HC15a shows that $\Sigma_{V_{\rm max}}$ is remarkably constant for a large range of 
$V_{\rm max}$ from the scales of dwarf to spiral/elliptical galaxies, namely 
$\Sigma_{V_{\rm max}}$ is confined within $10^1 \sim 
10^2$ $M_{\odot}$~pc$^{-2}$ over $10 \lsim V_{\rm max} \lsim 400$ km~s$^{-1}$. This 
implies that $\Sigma_{V_{\rm max}}$ is a common surface density scale for the 
description of a dark halo structure.

\begin{figure}[htpd]
\centering
\includegraphics[width=8.5cm]{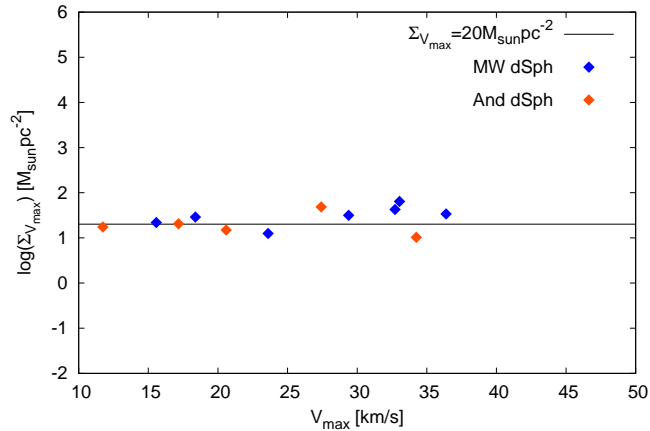}
\caption{
Mean surface density of a dark halo, $\Sigma_{V_{\rm max}}$, within a radius at 
maximum circular velocity $V_{\rm max}$, as a function of $V_{\rm max}$, for the 
dSphs in the Milky Way (red diamonds) and Andromeda (blue diamonds) taken from 
HC15a. Horizontal solid line denotes $\Sigma_{V_{\rm max}} = 
20$ $M_{\odot}$~pc$^{-2}$ (or $\log \Sigma_{V_{\rm max}} = 1.3$).}
\label{fig: Sigma_obs}
\end{figure}

For our current evolution model of dSphs, we confine ourselves to dwarf satellites in the 
Milky Way and Andromeda galaxies and adopt the fiducial value of $\Sigma_{V_{\rm 
max}}$ for this sample of dSphs. Figure \ref{fig: Sigma_obs} shows $\log 
\Sigma_{V_{\rm max}}$ vs. $V_{\rm max}$ for these dSphs taken from the HC15a 
paper. It is found that $\Sigma_{V_{\rm max}} = 20$ $M_{\odot}$~pc$^{-2}$ (or $\log 
\Sigma_{V_{\rm max}} = 1.3$) describes the common surface-density scale
in these dSphs quite well. In what follows, we use this value of
$\Sigma_{V_{\rm max}}$ as a fiducial common surface-density scale to describe
the evolution of dark halos in dwarf satellites, and further consider neighboring
values of $\Sigma_{V_{\rm max}} = 10$ $M_{\odot}$~pc$^{-2}$ (or $\log\Sigma_{V_{\rm max}} = 1.0$)
and 40 $M_{\odot}$~pc$^{-2}$ (or $\log\Sigma_{V_{\rm max}} = 1.6$) to investigate
the dependence of the following results on the variation of $\Sigma_{V_{\rm max}}$.

\subsection{Evolution of the NFW dark halos}
To obtain the evolution of dark halos under the constraint that $\Sigma_{V_{\rm max}} 
\simeq 20$ $M_{\odot}$~pc$^{-2}$, we here assume that the halo density profile follows the 
NFW profile with a central density $\rho_s$ and a scale length $r_s$, 
\begin{equation}
\rho_(r) = \frac{\rho_s}{(r/r_s)(1+r/r_s)^2} \ .
\label{eq: NFW}
\end{equation}
For this density profile, $V_{\rm max}$ is attained at $r_{\rm max}=2.16 r_s$ and 
$\Sigma_{V_{\rm max}}$ is then exactly proportional to $\rho_s r_s$. This NFW profile is 
parametrized by a concentration parameter $c = r_{\rm vir} / r_s$, where $r_{\rm 
vir}$ is the virial radius defined such that the mean density inside $r_{\rm vir}$ equals 
200 times the critical density of the Universe. We use the calibration of the halo virial 
mass, $M_{\rm vir}$,  and redshift dependence of the median halo concentration, 
$c(M_{\rm vir}, z)$, derived by \citet{Prada2012} for halos in the Bolshoi and 
MultiDark simulations, as adopted in \citet{Milosavljevic2014}.

\begin{figure}[htpd]
\centering
\includegraphics[width=8.5cm]{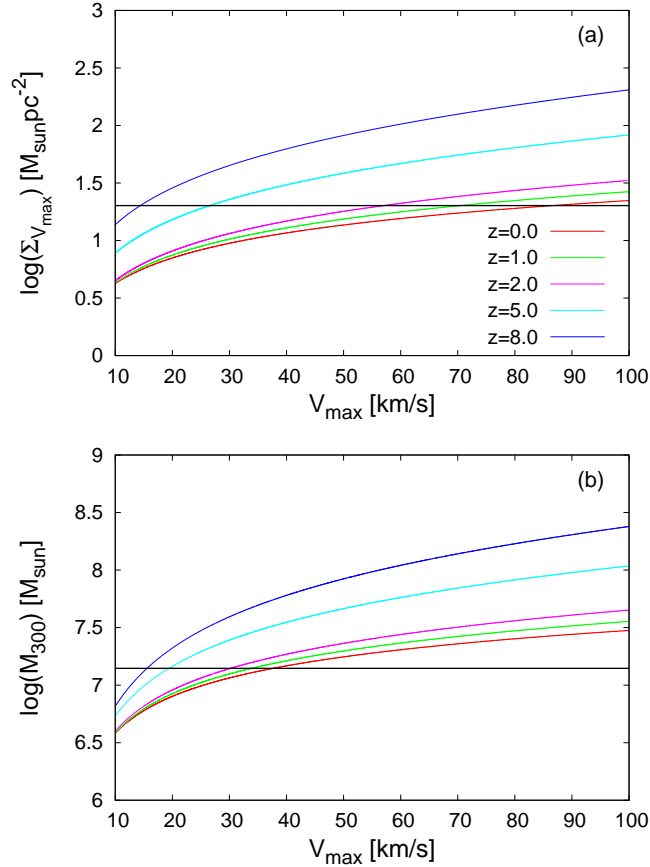}
\caption{
(a) Redshift evolution of $\Sigma_{V_{\rm max}}$ vs. $V_{\rm max}$ at $z=0$ for the 
NFW dark halo model. Horizontal line denotes $\Sigma_{V_{\rm max}} = 
20$ $M_{\odot}$~pc$^{-2}$ (or $\log \Sigma_{V_{\rm max}} = 1.3$).
(b) Redshift evolution of $M_{300}$ vs. $V_{\rm max}$ at $z=0$ for the NFW dark halo 
model. Horizontal line denotes $M_{300} = 1.4 \times 10^7$ $M_{\odot}$ obtained 
for classical dwarfs.}
\label{fig: NFW}
\end{figure}

Figure \ref{fig: NFW}(a) shows $\Sigma_{V_{\rm max}}$ vs. $V_{\rm max}$ at 
$z=0$ for this NFW dark halo model. As is clear from this figure, a mean surface density 
inside $r_{\rm max}$ shows only a weak dependence on the current maximum circular 
velocity or total mass of a dark halo. This reflects that such inner regions of all the halos 
having different total masses formed basically from very early mergers of small and dense 
halos and these high density regions have been rather undisturbed by later accretion of 
halos. Nonetheless, the mean surface density of halos is not exactly constant for 
different mass but varies, depending on redshifts, in comparison with the observed 
constancy of $\Sigma_{V_{\rm max}} = 20$ $M_{\odot}$~pc$^{-2}$ (horizontal black 
line).

This property of a halo is also seen when we adopt an inner mass of a dark halo instead 
of an inner surface density as a common scale;
Figure \ref{fig: NFW}(b) shows an inner mass, $M_{300}$, 
enclosed within $r=300$~pc \citep{Strigari2008} as a function of $V_{\rm max}$. As 
already demonstrated in \citet{Maccio2009} \citep[see also][]{Kravtsov2010}, $M_{300}$ also 
shows a weak dependence on $V_{\rm max}$ of a halo for the same reason given for
$\Sigma_{V_{\rm max}}$. In this 
panel, the horizontal black line corresponds to $M_{300} = 1.4 \times 
10^7$ $M_{\odot}$, which is a likely common value when we adopt the results of Jeans 
analysis only for bright, classical dwarfs under the assumption of spherical symmetry 
\citep{Strigari2008}; including ultra-faint dSphs (UFDs) suggests a somewhat smaller 
common scale of $M_{300} \simeq 10^7$ $M_{\odot}$, but UFDs generally show 
a large deviation from spherical symmetry and thus this latter value needs to be taken 
in caution. 

Comparing both panels in Fig. \ref{fig: NFW}, it follows that the common surface 
density scale, $\Sigma_{V_{\rm max}} = 20$ $M_{\odot}$~pc$^{-2}$, covers quite
a large range of $V_{\rm max}$ at different redshifts, whereas $M_{300} = 1.4 \times 
10^7$ $M_{\odot}$ is applicable only for $V_{\rm max} \lsim 40$ km~s$^{-1}$. 
This suggests that the use of $\Sigma_{V_{\rm max}}$ may be more general
in comparison with $M_{300}$ as a common scale for the evolution of a dark halo.

\begin{figure}[htpd]
\centering
\includegraphics[width=8.5cm]{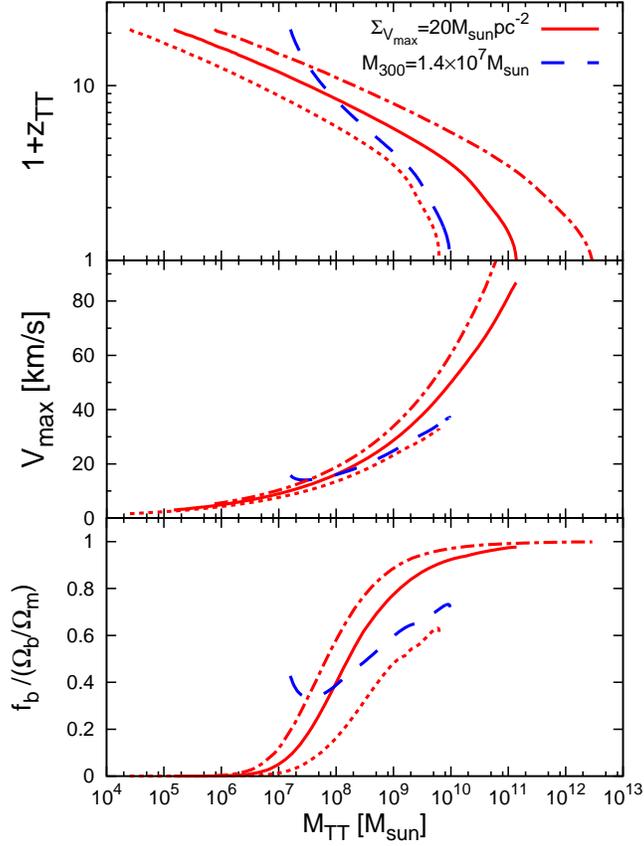}
\caption{
Properties of the dark halos with mass $M_{\rm TT}$ having the common 
surface-density scale of $\Sigma_{V_{\rm max}} = 20$ $M_{\odot}$~pc$^{-2}$ (solid 
line), compared with the case for the common mass scale of  $M_{300} = 
1.4\times10^7$ $M_{\odot}$ (dashed line).
The dot-dashed and dotted lines show, respectively, the cases of
$\Sigma_{V_{\rm max}} = 40$ $M_{\odot}$~pc$^{-2}$ and 10 $M_{\odot}$~pc$^{-2}$.
Top panel: redshift $z_{\rm TT}$ vs. 
$M_{\rm TT}$. Middle panel: maximum circular velocity $V_{\rm max}$ vs. $M_{\rm 
TT}$. Bottom panel: baryon fraction in units of the universal baryon fraction $f_b 
(\Omega_b /\Omega_m)^{-1}$ vs. $M_{\rm TT}$.}
\label{fig: TT}
\end{figure}

\subsection{Basic properties of dark halos with the common surface-density scale}
To understand the constancy of $\Sigma_{V_{\rm max}}$ at redshift $z=0$ and that of 
$M_{300}$ as well, we need to consider the termination of mass growth in these
dark halos associated with satellite galaxies, when they are accreted into a host 
halo, i.e. the halo of the Milky Way or Andromeda, due to tidal force. Following the 
model of Milosavljevi\'c \& Bromm (2014), we assume that the halos have a tidally 
truncated maximum mass, $M_{\rm TT}$, when they are accreted into a host halo
at redshift $z_{\rm TT}$. Then, we obtain a family of a dark halo with $M_{\rm TT}
(z_{\rm TT})$ so as to satisfy the common surface-density scale,
$\Sigma_{V_{\rm max}} = 20$ $M_{\odot}$~pc$^{-2}$. 

The top panel in Fig. \ref{fig: TT} shows $1 + z_{\rm TT}$ vs. $M_{\rm TT}$
under this common 
surface-density scale (solid line) in comparison with the common $M_{300}$ scale 
(dashed line). We also plot the cases of $\Sigma_{V_{\rm max}}
= 40$ $M_{\odot}$~pc$^{-2}$ (dot-dashed line) and 10 $M_{\odot}$~pc$^{-2}$ (dotted line)
for comparison. For the common surface-density scale of
$\Sigma_{V_{\rm max}} = 20$ $M_{\odot}$~pc$^{-2}$,
we obtain solutions with 
non-negative tidal truncation redshifts for the mass range of  $M_{\rm TT} < 2 \times 
10^{10}$ $M_{\odot}$, which is an order of magnitude wider than the case of the 
common $M_{300}$ scale ($M_{\rm TT} < 1.2 \times 10^{9}$ $M_{\odot}$).
This result reflects the property shown in Fig. \ref{fig: NFW} that
$\Sigma_{V_{\rm max}} = 20$ $M_{\odot}$~pc$^{-2}$ covers a larger range of $V_{\rm max}$
than $M_{300} = 1.4\times10^7$ $M_{\odot}$. While the 
increase of $M_{\rm TT}$ with decreasing $z_{\rm TT}$ is gradual at $z_{\rm TT} \le 
2$ on both scales, the slope is steeper at higher $z_{\rm TT}$, where the case of the 
common $\Sigma_{V_{\rm max}}$ scale provides a more rapid increase of $M_{\rm 
TT}$ with decreasing $z_{\rm TT}$ than the common $M_{300}$ scale. These 
properties of dark halo evolution are also seen in Fig \ref{fig: NFW}, where 
$z=const.$ lines are less dense at $z > 2$, especially for the evolution of 
$\Sigma_{V_{\rm max}}$ at smaller $V_{\rm max}$. This suggests that there is a 
systematic difference in the mass growth of a dark halo at high redshifts between the 
cases of the common $\Sigma_{V_{\rm max}}$ and $M_{300}$ scales.
We note that adopting the different value of $\Sigma_{V_{\rm max}} = 10$ or 40
$M_{\odot}$~pc$^{-2}$ leads to the change of an upper limit for $M_{\rm TT}$ by
an order of magnitude but the general dependence of $1 + z_{\rm TT}$ on $M_{\rm TT}$
remains basically the same. 

The middle panel in Fig. \ref{fig: TT} shows $V_{\rm max}$ for these common scales.
As already mentioned in Fig. \ref{fig: NFW}, a dark halo with $\Sigma_{V_{\rm max}} = 
20$ $M_{\odot}$~pc$^{-2}$ covers the large mass range of $10^5$ $M_{\odot}$ to 
$10^{11}$ $M_{\odot}$, so that the associated range for $V_{\rm max}$ is $5 \lsim 
V_{\rm max} \lsim 90$ km~s$^{-1}$, namely including both the small- and large-mass
galaxies. This is in contrast to the common 
$M_{300}$ scale, where the applicable range of $V_{\rm max}$ is limited. 
This limitation for the range of $V_{\rm max}$ is also seen when we adopt
$\Sigma_{V_{\rm max}} = 10$ $M_{\odot}$~pc$^{-2}$: only $V_{\rm max} \lsim 32$ km~s$^{-1}$
is allowed, which can also be deduced from Fig. \ref{fig: NFW}(a).

\section{Evolution of the baryonic component}
\subsection{Baryonic content}
Formation of dSphs is intimately affected by the cosmic reionization, whereby gas inflow 
into dark-matter subhalos associated with dSphs is inhibited and/or gas is flowing out of 
a gravitational potential due to photoionization. To what extent baryon is actually 
retained in each of subhalos is thus dependent on the evolution of photo-ionized gas in 
the presence of the ionizing background radiation, given the redshift evolution of dark 
halo density profiles on the scale of dSphs. This baryonic content is quantified 
by the baryon fraction $f_b$ in each mass of dark halos in units of universal baryonic 
fraction $\Omega_b / \Omega_m$, where $\Omega_b$ and $\Omega_m$ denote 
cosmic densities of baryon and matter normalized by a critical density of the Universe, 
respectively.

In this work, for the purpose of highlighting the effects of adopting different 
evolutionary histories of dark halos on the estimate of $f_b$, we follow the prescription 
for formulating it by Milosavljevi\'c \& Bromm (2014), using the critical halo mass, 
$M_{\rm crit}$, for retaining the half of universal baryon mass, defined as $f_b 
(M_{\rm crit}, z) = \Omega_b / 2\Omega_m$. We adopt $M_{\rm crit} (z)$ as given in 
\citet{Milosavljevic2014} and follow the dependence of $f_b$ on the halo mass 
$M_{\rm halo}$ given in \citet{Gnedin2000},
\begin{equation}
f_b (M_{\rm halo}, z) = \frac{\Omega_b}{\Omega_m}
 \left( 1 + (2^{\alpha/3} - 1) 
   \left[ \frac{M_{\rm halo}}{M_{\rm crit}(z)} \right]^{-\alpha} \right)^{-3/\alpha} 
\ ,
\end{equation}
where $\alpha$ is a shape parameter, for which we adopt $\alpha = 1$
\citep{Gnedin2012,Sobacchi2013} \citep[see also,][]{Okamoto2008}.
For the $z=0$ value of $M_{\rm crit}(z)$, we adopt $M_{\rm crit}(0)= 7\times10^9M_{\odot}$
\citep{Okamoto2008} as in \citet{Milosavljevic2014}.

The bottom panel in Fig. \ref{fig: TT} shows the baryon fraction in units of $\Omega_b / 
\Omega_m$ at the epoch of the tidal truncation of mass assembly.
It is found that for the low mass range of dark halos formed
at high $z_{\rm TT}$. their virial temperatures are too low to retain baryonic gas,
i.e., $f_b = 0$, on such low-mass scales. At 
lower redshifts of $z_{\rm TT} \lsim 10$, $f_b$ is increasing with $M_{\rm TT}$ and 
tending to the universal baryon fraction at the high-mass end. On the contrary, for the 
case of the common $M_{300}$ scale, since dark halos are already massive enough at high 
$z_{\rm TT}$ to retain baryon and their masses grow only gradually with decreasing 
$z_{\rm TT}$ as shown in the top panel of Fig. \ref{fig: TT},
the dependence of $f_b$ on $M_{\rm TT}$ is somewhat weaker than the case of
the common $\Sigma_{V_{\rm max}}$ scale.
This difference in the dependence of the baryon fraction on $M_{\rm TT}$ between
these common scales results in the different implications for star formation activity
in dwarf satellites as explained below.

\subsection{Comparison with observed dwarf satellites}
The estimated value of the baryon content $f_b$ at the epoch of the tidal truncation of
mass assembly is likely a maximum amount of available gas to form stars in each of
dwarf satellites. Then, during the resultant course of orbital motions within a
gravitational potential of a host halo, gas in each satellite may be lost in the form
of outflow driven by supernova feedback or eventually completely removed by ram
pressure stripping, at which star formation is finally quenched.

We estimate how much gas has actually been converted into stars in each of 
dwarf satellites using the observationally determined stellar mass and then
derive how much gas has been eventually lost from each system. We define
\begin{equation}
F = \frac{M_{\rm star}}{M_{\rm baryon}} \ , 
\label{eq: F}
\end{equation}
where $M_{\rm star}$ denotes the total mass of stars observed at the current epoch
and $M_{\rm baryon}$ is the available gas mass (estimated from $f_b$ at the epoch of
the tidal truncation of mass assembly) in each of satellite galaxies.
$F$ is thus a measure of star formation efficiency, where $F=1$ indicates that all the 
gas has been converted into stars and $F<1$ implies the loss of gas from the system by 
external effects such as ram-pressure and/or tidal stripping process or internal effects 
such as heating of gas from supernova feedback. Thus, $1-F$ tells us the total amount of 
gas lost from the system, thereby providing important information for understanding 
star formation and chemical evolution histories in dwarf galaxies.

\begin{figure}[htpd]
\centering
\includegraphics[width=9cm]{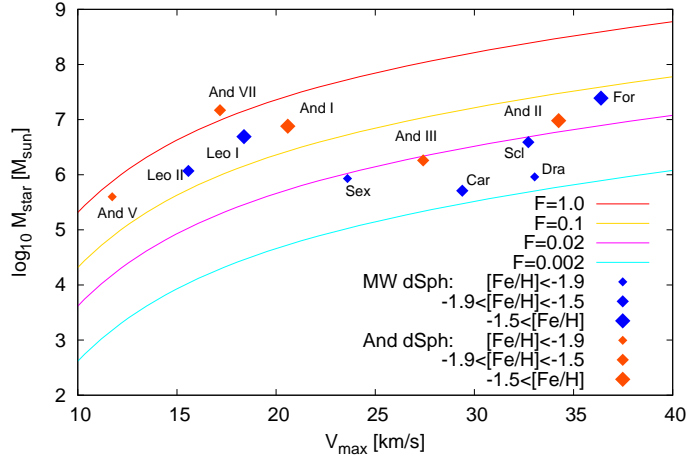}
\caption{
Stellar mass $M_{\rm star}$ vs. $V_{\rm max}$ for the dSphs in the Milky Way
(red diamonds) and Andromeda (blue diamonds), with larger sizes of the marks
for higher metallicities.
Solid lines shows, from top to bottom, the semi-analytical model for
$F=1.0$, 0.1, 0.02 and 0.002
in the fiducial case of
$\Sigma_{V_{\rm max}} = 20$ $M_{\odot}$~pc$^{-2}$.}
\label{fig: F}
\end{figure}

Figure \ref{fig: F} shows $M_{\rm star}$ vs. $V_{\rm max}$ for dSphs in the Milky Way and 
Andromeda galaxies. The mass of stars in each dSph is based on the estimate of the
mass-to-luminosity ratios in the $V$-band magnitude, $M_{\rm star} / L_{\rm V}$, by \citet{Woo2008},
where the values of $L_{\rm V}$ for these sample galaxies are taken from the data assembled and
listed in \citet{Kirby2014}. Their mean metallicities are shown in terms of 
the size of each mark, for which the data are taken from \citet{Kirby2013}.
Solid lines show our semi-analytical model for constant $F$ sequences,
in the fiducial case of $\Sigma_{V_{\rm max}} = 20$ $M_{\odot}$~pc$^{-2}$.
It follows 
from the figure that the current sample of dwarf satellites is divided into roughly two 
different families of $F > 0.1$ at $V_{\rm max} \lsim 22$ km~s$^{-1}$ (And~V, Leo~II, And~VII,
Leo~I and And~I) and $F < 0.1$ at $V_{\rm max} \gsim 22$ km~s$^{-1}$ (Sextans, And~III, Carina,
Sculptor, Draco, And~II and Fornax). The former, low mass dSphs, are characterized by
$z_{\rm TT} \gsim 6$, at which stars formed efficiently before the
effect of the reionization became significant, whereas star formation efficiency in the latter,
more massive dSphs may be reduced because of the presence of the ionizing background by
the time they are accreted into a host halo. It is also suggested that except for 
And~VII, all the dSphs imply $F < 0.3$ and thus more than 70~\% of gas is lost from the system. 
This is indeed supported by recent chemical evolution models \citep[e.g.,][]{Homma2015} 
to understand the observed metal-poor distributions of stars in dSphs.

For And~VII, we obtain $F=1.45$ with $M_{\rm baryon} = 1.02\times 10^7$ $M_{\odot}$
at $z_{\rm TT} = 7.06$ and $M_{\rm star}=1.48\times 10^7$ $M_{\odot}$, suggesting quite an
efficient conversion of gas into stars, although $M_{\rm star} > M_{\rm baryon}$
at the redshift of the tidal truncation needs explanations. This is 
consistent with the {\it Hubble Space Telescope} results by \citet{Weisz2014} that more than
90~\% of the stars in And~VII formed within the first 1~Gyr. Regarding $F>1$ in this galaxy,
it is interesting to note that the distance from the Andromeda center to this satellite is
largest (218 kpc) among the current sample of the Andromeda satellites, whereby this galaxy
may be allowed to obtain the further gas infall after the epoch of $z_{\rm TT}$ 
and form extra stars before the ram pressure stripping removes gas completely.

\begin{figure}[htpd]
\centering
\includegraphics[width=9cm]{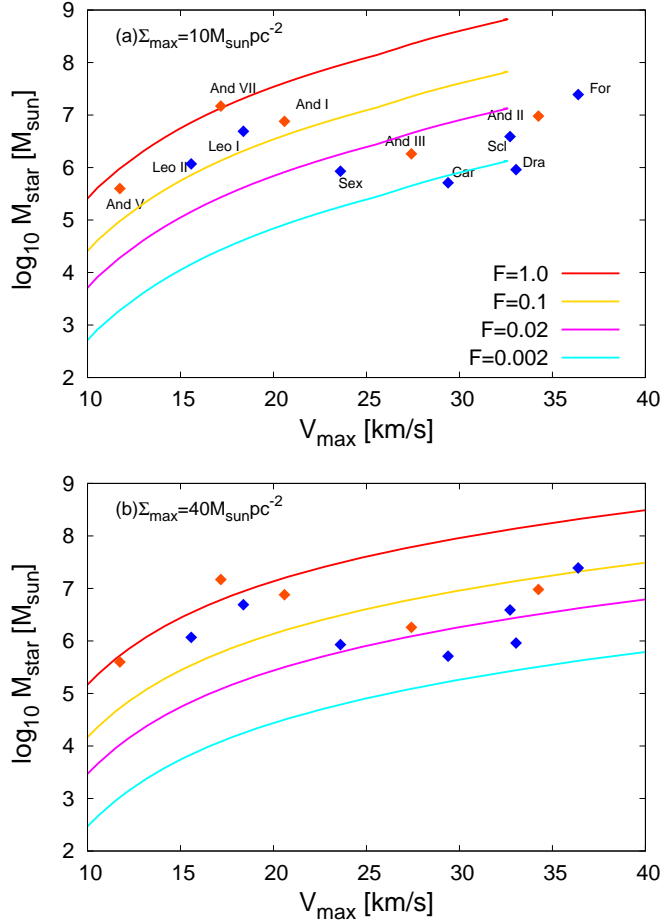}
\caption{
The same as Fig. \ref{fig: F}, but for the different common surface-density scales of
$\Sigma_{V_{\rm max}} = 10$ $M_{\odot}$~pc$^{-2}$ (a) and
40 $M_{\odot}$~pc$^{-2}$ (b). For the former case, solutions are available
only for $V_{\rm max} \le 32$ km~s$^{-1}$, as shown in the middle panel
of Fig. \ref{fig: TT}.}
\label{fig: F2}
\end{figure}

Figure \ref{fig: F2} shows the effects of adopting neighboring values
of $\Sigma_{V_{\rm max}} = 10$ $M_{\odot}$~pc$^{-2}$ (a) and 40 $M_{\odot}$~pc$^{-2}$ (b).
It follows that the general properties of constant $F$ lines relative to observed dSphs
remain basically the same as the fiducial case of $\Sigma_{V_{\rm max}} = 20$ $M_{\odot}$~pc$^{-2}$,
namely the dSphs can be divided into those with high or low $F$, depending on the value of
$V_{\rm max}$. In addition, compared with the fiducial case of
$\Sigma_{V_{\rm max}} = 20$ $M_{\odot}$~pc$^{-2}$, adopting a smaller (larger)
$\Sigma_{V_{\rm max}}$ results in a smaller (larger) $F$, thereby implying that
dSphs having higher surface densities at a given $V_{\rm max}$ are expected to show
higher star formation efficiency and thus may cause some dispersion in the distribution of $F$.

We note that in the \citet{Milosavljevic2014} paper adopting $M_{300}=const.$,
halos with high baryon fraction are thought to form stars in isolated/nuclear clusters,
whereas halos with low baryon fraction imply low star formation efficiency and hence
galaxies are expected to stay dark. Thus, the adoption of a different common scale
leads to a different interpretation for star formation activity in dSphs.

\section{Discussion}
\subsection{Comparison with the high-resolution numerical simulation}
Our model of dwarf galaxies in the common surface-density scale combined with gas 
physics in the ionizing background suggests that there are broadly two families of satellite 
galaxies, those experiencing efficient star formation with $F > 0.1$ at early epochs 
and those with inefficient star formation with $F < 0.1$ due to the effects of the ionizing 
background at lower redshifts.
These different types of satellite galaxies may be divided at $V_{\rm 
max}$ of about 22 km~s$^{-1}$. However, the number of sample dSphs in the Milky 
Way and Andromeda galaxies for the current analysis is limited only to 11 at a moment,
so it is yet unclear whether this suggested property is indeed the case.

\begin{figure}[htpd]
\centering
\includegraphics[width=9cm]{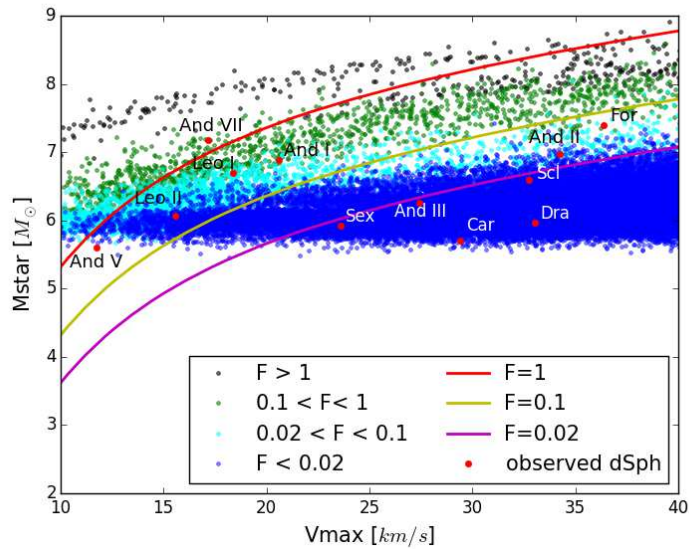}
\caption{
$M_{\rm star}$ vs. $V_{\rm max}$ for subhalos in the Illustris simulation (dots). Simulation 
results are divided into 4 different ranges in $F$. Solid lines correspond to our 
semi-analytical models for $F$ and observed data are shown with red circles.}
\label{fig: simulation}
\end{figure}

To remedy this situation, we investigate the results of the recent high-resolution
numerical simulation for galaxy formation in the fully cosmological context,
where the reionization of the Universe is taken into account. For this purpose,
we adopt the Illustris simulation \citep{Nelson2015}. We derive $F$ defined in
eq. (\ref{eq: F}) associated with each dark-matter subhalo at $z=0$ given in
the Illustris data, where $M_{\rm baryon}$ is estimated by multiplying the mass
of each subhalo by the universal baryon fraction of $\Omega_{b}/\Omega_{m}$.
Note that in this procedure of deriving $M_{\rm baryon}$ from the simulation results,
a finite fraction of subhalos may have continued their growth up to $z=0$, in
contrast to our semi-analytical model assuming the tidal truncation at $z_{\rm TT}$.
This suggests that we may overestimate $M_{\rm baryon}$ and thus underestimate
$F$ from the simulation results to some extent, compared with those
in our semi-analytical model.

Figure \ref{fig: simulation} shows the $M_{\rm star}$ vs. $V_{\rm max}$ for
the simulated subhalos, which are divided into 4 different 
ranges in $F$. Solid lines correspond to our semi-analytical models for $F$ and 
observed data are shown with red circles. Although there exists some discrepancy 
between our model and numerical simulation results at $V_{\rm max} < 
20$ km~s$^{-1}$, which may be partially due to insufficient resolution in
numerical simulation on such small scales or other effects such as later
gas infall into subhalo after the tidal truncation by their host halo,
it is quite impressive that
a very good agreement is achieved at $V_{\rm max} > 25$ km~s$^{-1}$,
including the slope of a constant $F$ line in the diagram, 
although we never make any fine tuning in our model.

\begin{figure}[htpd]
\centering
\includegraphics[width=9cm]{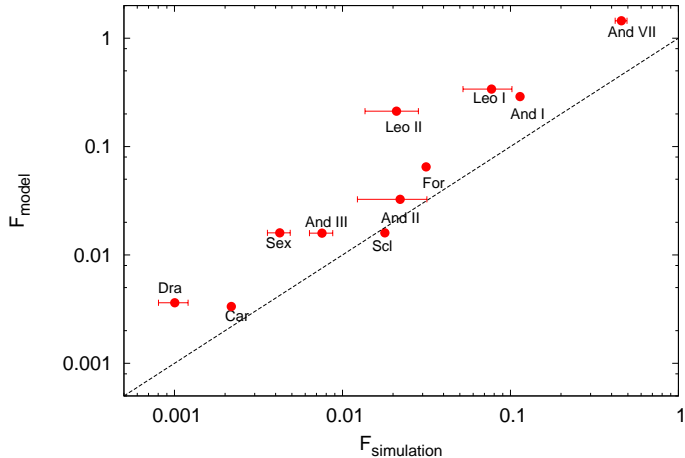}
\caption{
Comparison between $F$ obtained from the Illustris simulation and that from our 
semi-analytical model.}
\label{fig: F_comp}
\end{figure}

Indeed, Fig. \ref{fig: F_comp} shows the comparison between $F$ obtained from the 
Illustris simulation and that from our semi-analytical model. This figure clearly 
demonstrates that there is a very reasonable agreement between these values of $F$,
although $F$ from the simulation results appears slightly smaller than that in the semi-analytical
model due to the reason mentioned above. This may support our hypothesis 
that there exist roughly two different families of dwarf galaxies in terms of star formation 
efficiency, which is probably controlled by external and/or internal feedback effects 
including the ionizing background radiation and/or supernova explosions. 

\begin{figure}[htpd]
\centering
\includegraphics[width=9cm]{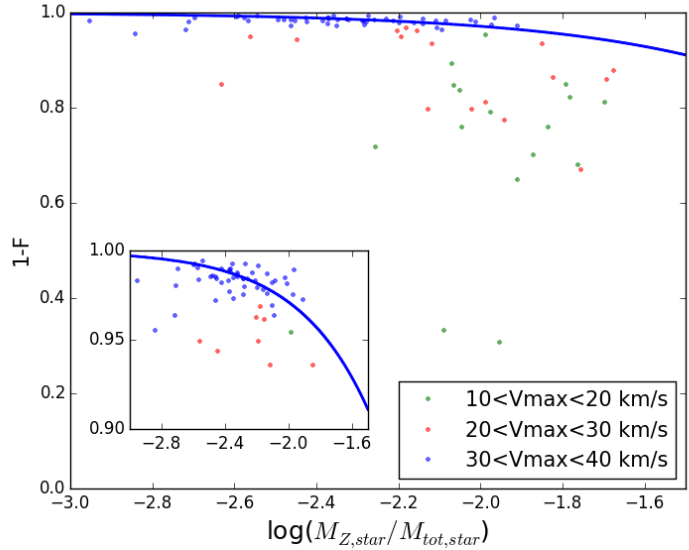}
\caption{
Comparison between $1-F$ and the stellar metallicity in each subhalo for the Illustris simulation,
where the latter is estimated from the mass-weighted average metallicity of the star particles
$M_{\rm Z,star}/M_{\rm tot,star}$. In this plot, to avoid the effect of the mass-metallicity
degeneracy, we adopt only the subhalos having
$M_{\rm tot,star} = (1.00 \pm 0.05) \times 10^7$ $M_{\odot}$.
Solid line denotes the prediction of a fiducial chemical
evolution model including a gaseous outflow, where the outflow rate is assumed to be proportional
to the star formation rate and we adopt the resultant effective yield as 0.7.
The inset shows the plots in the range of $0.9 \le 1-F \le 1.0$ for clarity.
}
\label{fig: 1-F}
\end{figure}

With this $F$ value, it is also possible to estimate the integrated amount of mass loss
from dwarf satellites in terms of $1-F$, which may ultimately affect their chemical
evolution histories. To infer this property in the Illustris simulation data, we
plot, in Fig. \ref{fig: 1-F}, $1-F$ vs. the mass-weighted average metallicity of the
star particles $M_{\rm Z,star}/M_{\rm tot,star}$ in each subhalo, for different ranges of
$V_{\rm max}$. In this plot, to avoid the effect of the mass-metallicity
degeneracy in the result, we adopt only the subhalos having
$M_{\rm tot,star} = (1.00 \pm 0.05) \times 10^7$ $M_{\odot}$.
Although there are large dispersions in the plot, it follows that a subhalo with
larger $1-F$ tends to have a more metal-poor stellar system. 
The solid line corresponds to a fiducial chemical evolution model including a gaseous outflow,
where the outflow rate is assumed to be proportional to the star formation rate and
the resultant effective yield is taken as 0.7. Although we do not intend to fit this line
to the simulation data, the model reasonably reproduces the properties of subhalos
especially with $V_{\rm max}$ larger than $\sim 30$ km~s$^{-1}$,
thereby implying that a subhalo losing more gas and metals as well leaves
a more metal-poor dwarf galaxy. We note that
for subhalos with smaller $V_{\rm max}$, the simulation data show systematically
more metal-poor stars for given $1-F$ than the case of the solid line.
Physically this may imply
the loss of metals independent of the loss of gas in such subhalos,
e.g. by means of the metal-enhanced supernova wind
\citep{MacLow1999,Martin2002,Kirby2011a,Kirby2011b,McQuinn2015}.
Alternatively, this may be just an artifact of rather low numerical resolution of
Illustris at $V_{\rm max} < 25$ km~s$^{-1}$, where as shown in Fig. \ref{fig: simulation},
our semi-analytical model for a constant $F$ differs from $F$ deduced from Illustris:
the latter predicts a systematically smaller $F$ than our model and thus
lower star formation efficiency, so that a stellar system is made more metal-poor
than the case for larger $V_{\rm max}$. Further investigation for this issue is needed,
based on much higher-resolution numerical simulations.

\subsection{Star formation history}
Dark halos in dwarf satellite galaxies continue to grow via mass accretion from outside 
until they enter a host halo at $z_{\rm TT}$ after which the resultant tidal stripping 
inhibits a further mass growth. The ram-pressure stripping may also be at work to 
prevent further gas infall and thus star formation. Thus we expect that the main star formation 
activity in each of dwarf galaxies occurs prior to $z_{\rm TT}$, where the continuous 
growth of dark halos accompanies gas inflow simultaneously, from which new
stars are formed.

Following this conjecture, we calculate the inflow rate of gas from outside, based on the 
associated mass growth in a dark halo component. For this purpose, we adopt the 
growth history of a dark halo, $d \ln M_{\rm halo} / dz$, given in equation (7) of the 
\citet{Milosavljevic2014} paper:
\begin{equation}
\frac{d\ln M_{\rm halo}}{dz} = -0.62 \left( \frac{1+1.11 z}{1+z} \right)
\left[ \frac{d\sigma^2/d\ln M(M_{\rm halo})}{d\sigma^2/d\ln M(10^{12}M_{\odot})} \right]^{-1/2}
\ , 
\label{eq: growth rate}
\end{equation}
where $\sigma(M)$ denotes the rms fluctuation of the density field linearly extrapolated to
$z=0$. 
This equation corresponds to the growth rate of the
mean most massive progenitor for halos, which now belong to the common surface-density
scale family in this work. We assume that the inflow rate of gas is simply given as
the baryon fraction times the growth rate of a dark halo, i.e., being proportional to
$dM_{\rm halo} / dt$.

\begin{figure}[htpd]
\centering
\includegraphics[width=10cm]{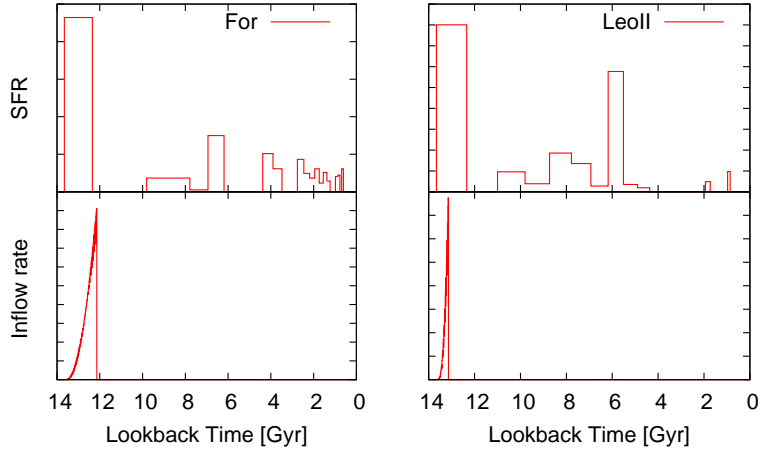}
\caption{
Upper panels: SFR at each epoch calculated from the observed cumulative SFR \citep{Weisz2014} 
for Fornax (left) and Leo~II (right). Lower panels: Gaseous infall rate based 
on our dark-halo model in the common surface-density scale for Fornax (left) and Leo~II 
(right). In both upper and lower panels, the vertical axis is arbitrarily normalized.}
\label{fig: inflow}
\end{figure}

In this work, we choose Fornax and Leo II as two representative cases,
where the epochs of tidal truncation under a constraint of the common surface-density scale
of 20 $M_{\odot}$~pc$^{-2}$ are given as $z_{\rm TT} = 3.75$ and $7.57$, respectively,
i.e. subsequent and prior to the completion of the cosmic reionization. For these
dSphs, the values of $F$ indicating star formation efficiency are estimated as 0.065 (Fornax)
and 0.212 (Leo II), thereby suggesting that the former dSph with $z_{\rm TT}$ being lower
than the redshift for the completion of the cosmic reionization is subject to the suppression
of star formation caused by the ionizing background radiation.

To estimate the star formation histories in these dSphs,
we adopt the {\it Hubble Space Telescope} results by \citet{Weisz2014}
for the Milky Way dSphs, based on the 
distribution of resolved member stars in the color-magnitude diagram. Since in the 
\citet{Weisz2014} paper, they showed the cumulative star formation rate (SFR), we 
evaluate the SFR at each epoch by differentiating the cumulative SFR. 
Upper panels in 
Fig. \ref{fig: inflow} show the (normalized) SFR at each epoch for Fornax (upper left) and Leo~II 
(upper right). Lower panels in this figure show our model for the (normalized) infall rate
of gas based on the dark matter growth in the common surface-density scale, for Fornax (lower left) 
and Leo~II (lower right). In these lower panels, the inflow rate of gas is maximum at 
redshift $z_{\rm TT}$, i.e. at the epoch of tidal truncation when a subhalo enters into a host halo 
at the corresponding epoch. Figure \ref{fig: inflow} suggests that the observed peak of the 
star formation rate in Fornax and Leo~II is indeed in good agreement with the epoch 
when the peak of the gas infall rate is achieved. This implies that star formation activity 
in these dwarf satellites may be controlled by the mass growth in a dark halo component
under a constraint of the common surface-density scale of 
$\Sigma_{V_{\rm max}} \simeq 20$ $M_{\odot}$~pc$^{-2}$ yielding $z_{\rm TT}$, i.e., 
the peak epoch of the gas infall rate, and this epoch is to be compared with
the completion epoch of the cosmic reionization to infer the effect on star formation efficiency.
Further tests for the current semi-analytical model are needed, in which
much tighter limits on star formation histories of these galaxies, e.g. based on
measurements of detailed abundance patterns of member stars, may enable us
to deduce any differences in star formation efficiency between the two families
of dSphs.

\section{Conclusions}
We have investigated the mass growth of dark halos in dwarf galaxy satellites in Local 
Group galaxies, under the recently proposed property by HC15a that the mean surface 
density of a dark halo inside a radius at maximum circular velocity is universal over a 
large range of luminosities and masses of galaxies. Following that this surface density 
$\Sigma_{V_{\rm max}}$ with 20 $M_{\odot}$~pc$^{-2}$ well explains dwarf satellite 
galaxies in the Milky Way and Andromeda galaxies, we calculate the evolution of dark 
halos and associated evolutionary histories of the baryonic component in dwarf satellites. 
Our main results are summarized as follows.

\begin{itemize}
\item
Dark halos in the common surface-density scale show the rapid mass growth at high 
redshifts, while they evolve only slowly at lower redshifts of $z_{\rm TT} < 2$. It is 
found that these dark halos cover a large range of $V_{\rm max}$ including the scales of 
UFDs with $V_{\rm max}$ as small as 10~km~s$^{-1}$ as well as those of bright 
galaxies with $V_{\rm max}$ as large as a few hundred km~s$^{-1}$. This is in 
contrast to dark halos in the common mass scale of $M_{300} \simeq 1.4 \times 
10^7$ $M_{\odot}$~pc$^{-2}$, which covers only a limited range of $V_{\rm max}$. 
Thus, the mean surface density $\Sigma_{V_{\rm max}}$ provides a more general
description for the evolution of dark halos.

\item
The calculation of the baryon fraction retained in these dark halos in the
presence of the ionizing background radiation reveals that low-mass satellites with $V_{\rm 
max}$ below $\sim 22$ km~s$^{-1}$ are characterized by high star formation efficiency 
of $F > 0.1$, whereas more massive satellites with $V_{\rm max}$ being larger than 
$\sim 22$ km~s$^{-1}$ show low star formation efficiency of $F < 0.1$. This is 
understood on the basis of the mass growth of dark halos and its termination at 
$z_{\rm TT}$ in the following manner. Low $V_{\rm max}$ halos grow fast at high 
redshifts and are less affected by the ionizing radiation, whereby star formation 
efficiency is high. On the contrary, high $V_{\rm max}$ halos grow slowly at lower 
redshifts, so that star formation is suppressed by the ionizing background. It is also 
suggested from these models that more than 70~\% of baryon (i.e., $1-F > 0.7$)
have been lost from each dwarf satellite, in good agreement with suggestions
from chemical evolution models to explain low metallicities of stars in the observed dSphs.

\item
We have found that the values of $F$ in subhalos calculated from the Illustris simulation are
in agreement with those derived from our semi-analytical models, implying that
our models may describe the realistic evolutionary histories of dark halos and baryonic matter,
although further studies based on much higher-resolution numerical simulations are needed
to assess the detailed properties of star formation activities in satellites, especially at
small $V_{\rm max}$.

\item
The infall rate of gas inferred from the growth rate of dark halos in the common surface-density 
scale shows a peak at the epoch of tidal truncation and this peak epoch is reasonably in
agreement with the epoch at which the star formation rate derived for the Milky Way satellites
(Fornax and Leo~II) is maximum. This also implies that the evolution of a dark halo may play
a key role in understanding star formation histories in dwarf satellite galaxies.

\end{itemize}

\acknowledgments
We are grateful to the referee for his/her invaluable comments on the manuscript.
We also thank Kohei Hayashi for useful discussion, which actually promotes 
the current work, and Evan Kirby for his constructive comments on
an early draft of this paper, which help us improve the manuscript.
This work is supported in part by JSPS Grant-in-Aid for Scientific 
Research (B) (No. 25287062) ``Probing the origin of primordial minihalos via 
gravitational lensing phenomena'' and MEXT Grant-in-Aid for Scientific Research on 
Innovative Areas ``Cosmic Acceleration'' (No. 15H05889).


\end{document}